\documentclass[apjl]{emulateapj_arxiv}
\usepackage{graphicx}

\usepackage{color}
\newcommand{\fig}[1]{Figure~\ref{#1}} 
\newcommand{\figs}[1]{Figures~\ref{#1}} 
\newcommand{\sref}[1]{Section~\ref{#1}} 
\newcommand{\eqref}[1]{Equation~\ref{#1}} 
\newcommand{\kms}{\,km\,s$^{-1}$}
\newcommand{\carcsec}{$\mbox{.\hspace{-0.5ex}}^{\prime\prime}$}

\definecolor{remgray}{rgb}{0.5,0.5,0.5}

\newcommand{\fei}{Fe\,\textsc{i}}
\newcommand{\coi}{Co\,\textsc{i}}

\newcommand{\imax}{{IMaX}}
\newcommand{\sufi}{\textsc{SuFI}}
\newcommand{\sunrise}{\textsc{Sunrise}}

\shortauthors{Lagg et al.}
\shorttitle{Resolved magnetic flux tube}
\begin{document}
\title{Fully resolved quiet-Sun magnetic flux tube observed with the
  \sunrise{} / \imax{} instrument}

\author{A.~Lagg\altaffilmark{1}, S.~K.~Solanki\altaffilmark{1,7},
  T.~L.~Riethm\"uller\altaffilmark{1}, V.~Mart\'inez Pillet\altaffilmark{2},
  M.~Sch\"ussler\altaffilmark{1}, J.~Hirzberger\altaffilmark{1},
  A.~Feller\altaffilmark{1}, J.~M.~Borrero\altaffilmark{3,1},
  W.~Schmidt\altaffilmark{3}, J.~C.~del~Toro~Iniesta\altaffilmark{4},
  J.~A.~Bonet\altaffilmark{2}, P.~Barthol,\altaffilmark{1},
  T.~Berkefeld\altaffilmark{3}, V.~Domingo\altaffilmark{5},
  A.~Gandorfer\altaffilmark{1}, M.~Kn\"olker\altaffilmark{6}, and
  A.~M.~Title\altaffilmark{8}}

\affil{$^1$Max-Planck-Institut f\"ur Sonnensystemforschung,
  Max-Planck-Stra\ss{}e 2, 37191 Katlenburg-Lindau, Germany}

\affil{$^2$Instituto de Astrof\'{\i}sica de Canarias, C/Via L\'actea s/n,
  38200 La Laguna, Tenerife, Spain}

\affil{$^3$Kiepenheuer-Institut f\"ur Sonnenphysik, Sch\"oneckstra\ss{}e 6,
  79104 Freiburg, Germany}

\affil{$^4$Instituto de Astrof\'{\i}sica de Andaluc\'{\i}a (CSIC), Apartado de
  Correos 3004, 18080 Granada, Spain}

\affil{$^5$Grupo de Astronom\'{\i}a y Ciencias del Espacio, Universidad de
  Valencia, 46980 Paterna, Valencia, Spain}

\affil{$^6$High Altitude Observatory, National Center for Atmospheric
  Research, P.O. Box 3000, Boulder, CO 80307-3000, USA}

\affil{$^7$School of Space Research, Kyung Hee University, Yongin, Gyeonggi
  446-701, Republic of Korea}

\affil{$^8$Lockheed Martin Solar and Astrophysics Laboratory, Bldg. 252, 3251
  Hanover Street, Palo Alto, CA 94304, USA}

\email{lagg@mps.mpg.de}




\begin{abstract}
  Until today, the small size of magnetic elements in quiet Sun areas has
  required the application of indirect methods, such as the line-ratio
  technique or multi-component inversions, to infer their physical
  properties. A consistent match to the observed Stokes profiles could only be
  obtained by introducing a magnetic filling factor that specifies the
  fraction of the observed pixel filled with magnetic field. Here, we
  investigate the properties of a small magnetic patch in the quiet Sun
  observed with the \imax{} magnetograph on board the balloon-borne telescope
  \sunrise{} with unprecedented spatial resolution and low instrumental stray
  light. We apply an inversion technique based on the numerical solution of
  the radiative transfer equation to retrieve the temperature stratification
  and the field strength in the magnetic patch. The observations can be well
  reproduced with a one-component, fully magnetized atmosphere with a field
  strength exceeding 1\,kG and a significantly enhanced temperature in the
  mid- to upper photosphere with respect to its surroundings, consistent with
  semi-empirical flux tube models for plage regions. We therefore conclude
  that, within the framework of a simple atmospheric model, the \imax{}
  measurements resolve the observed quiet-Sun flux tube.
\end{abstract}


%

\section{Introduction}

Ever since the 1970s \citep[e.g.][]{howard:72,frazier:72} it has been known
that small-scale magnetic elements are unresolved. Therefore, the magnetic
field strength averaged over the resolution element was found to be
significantly lower than the field strength of the magnetic structure. Their
intrinsic properties (field strength and temperature structure) could only be
obtained by introducing a magnetic filling factor to account for the fact that
only a certain fraction of the observed resolution element contains the
magnetic structure. This factor is sometimes also called a stray light factor,
although the physical meaning is somewhat different.  Indirect methods
\citep[see][for an overview]{solanki:93c} such as the line-ratio technique
\citep{stenflo:73} or the inversion of line profiles
\citep[e.g.][]{martinezpillet:97} pointed to the kilo-Gauss nature of the
small-scale magnetic elements in the network. The reality of these kilo-Gauss
field strengths has been confirmed by direct splitting of infrared lines
\citep{harvey:75,harvey:77b,rabin:92,ruedi:92b}. By employing speckle imaging
techniques, \citet{keller:92} observed Stokes $V$ amplitudes in plage regions
close to the values predicted by semi-empirical flux tube models. This
groundbreaking analysis suffers from the assumption that the chosen empirical
atmosphere is applicable to the particular observed feature.

Another basic property of small magnetic elements is that they are hotter than
their surroundings in the middle and upper photosphere, with intrinsic
temperature differences of $\approx$1000\,K being reached at $\log\tau\approx
-2$ \citep{solanki:86}. These high temperatures lead to a weakening of many
Fraunhofer lines, resulting in line gaps or discontinuities on spectrograms
\citep{sheeley:67}.

Here, we report on the self-consistent determination of the magnetic field
strength and temperature in the magnetic elements, without taking recourse to
a magnetic filling factor.  Whether the observed magnetic elements have
finally been resolved can then be judged by comparing with both the field
strengths and the excess temperature in magnetic elements given in the
literature as deduced with methods that allow for the magnetic filling factor.

\section{Observations}

\begin{figure}
  \centering
  \includegraphics[width=\linewidth]{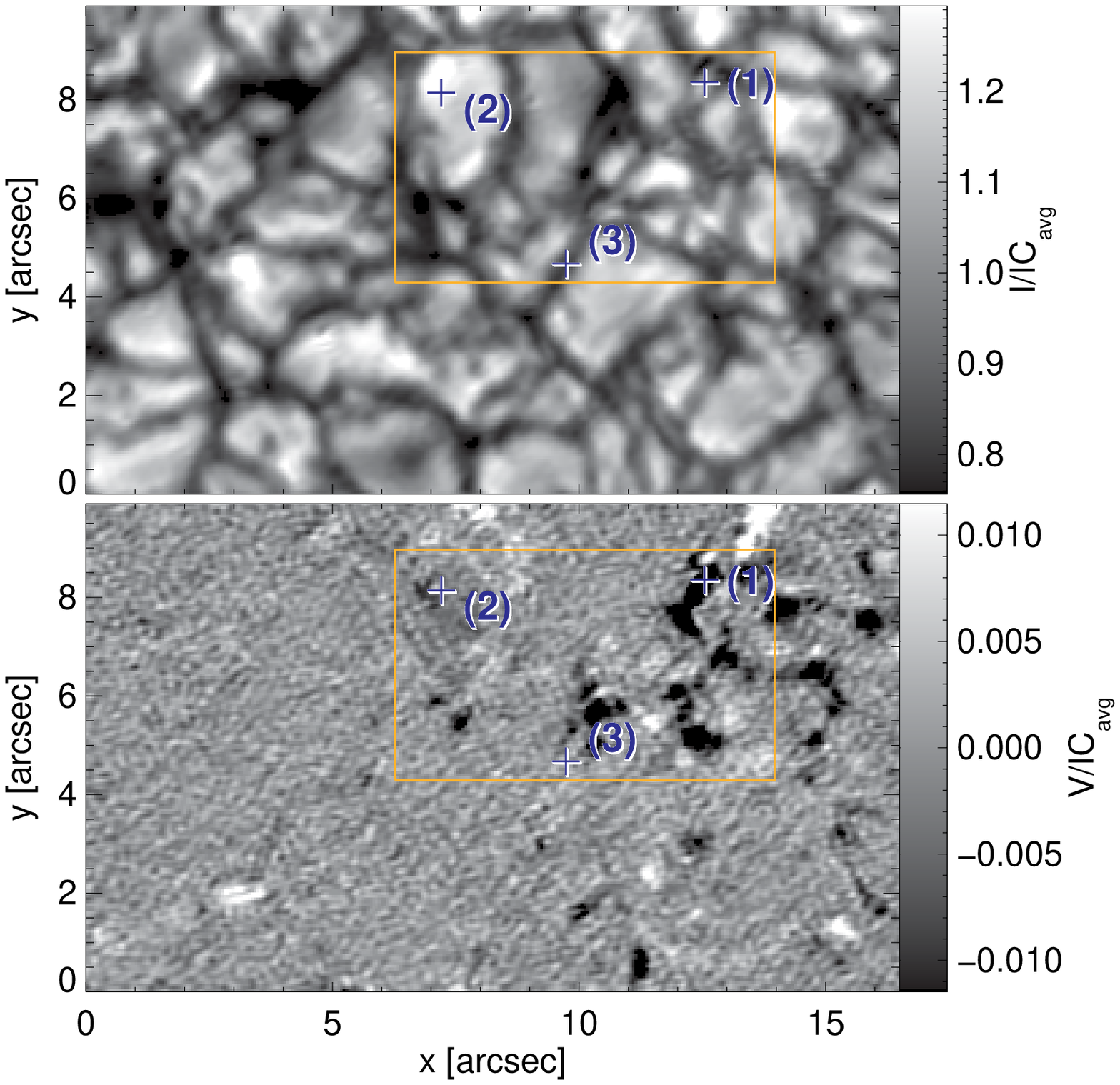}
  \caption{\label{fig1.eps}\imax{} continuum intensity map (top) and Stokes
    $V$ map (bottom) normalized to the continuum level of the average $I$
    profile (see \fig{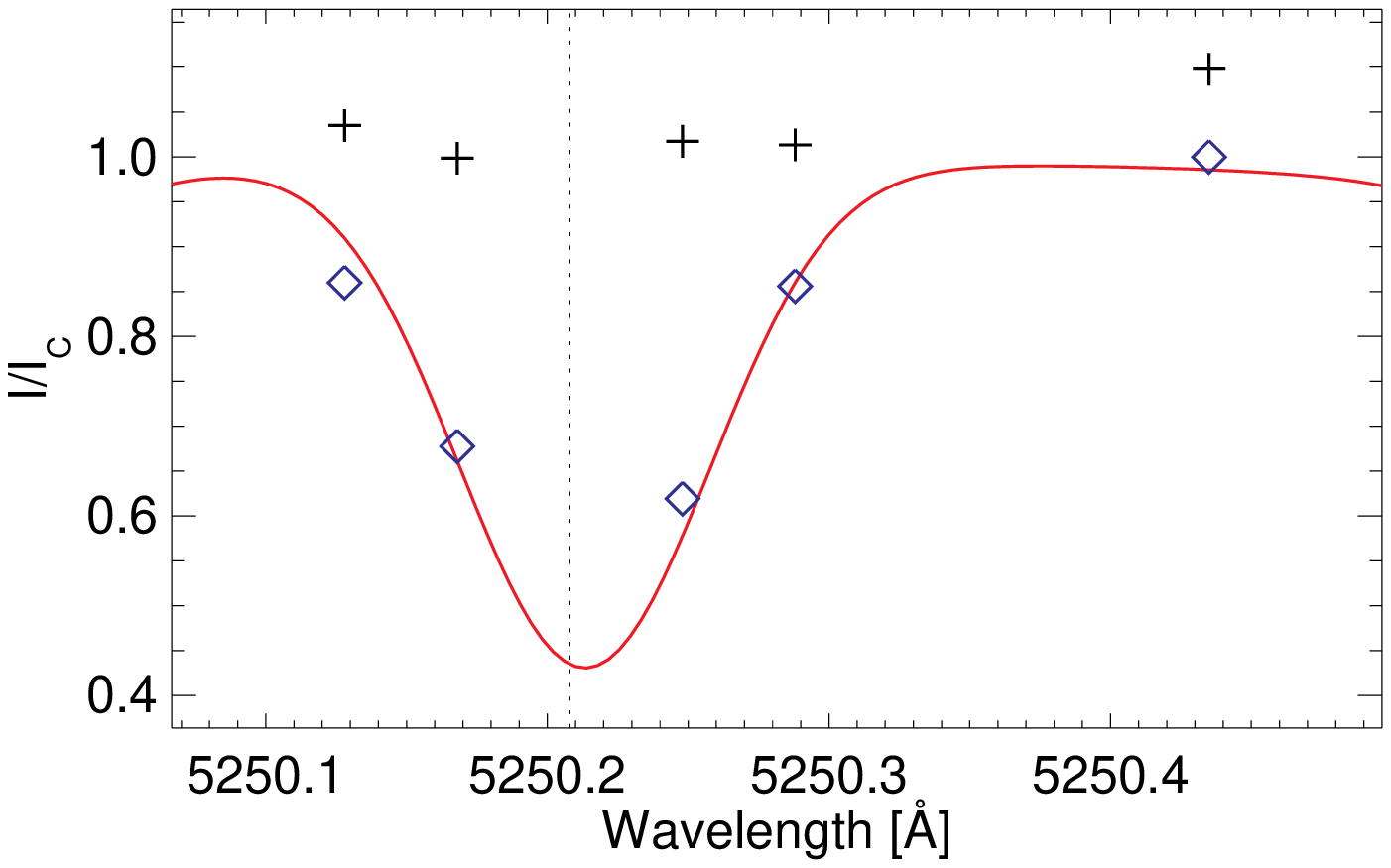}). The box encloses the region for the
    inversion maps in \sref{results}.}
\end{figure}

We use data from the Imaging Magnetograph eXperiment
\citep[\imax{},][]{martinezpillet:10a} on board the \sunrise{} balloon mission
\citep{barthol:10a}. For an overview of the \sunrise{} data set and a
description of selected results we refer to \citet{solanki:10a}. \imax{}
measured the full Stokes vector in five wavelength positions located at
($-80,-40,+40,+80,$ and $+227$)~m\AA{} from the line center of the
magnetically sensitive \fei{} 5250.2\,\AA{} line (Land\'e factor $g=3$). The
data set was obtained on 2009 June 9, at 00:36:45~UT (data set 163--209). The
observed region was located close to the disk center. The total field of view
of \imax{} covered 50\arcsec$\times$50\arcsec{} and contained a quiet-Sun area
with isolated network elements, visible in the Stokes $V$ map. The Stokes $V$
map was obtained by integrating over the two blue wavelength points ($-80$ and
$-40$\,m\AA{} from the line center). For the analysis in this Letter we discuss
in detail a 7\carcsec{}7 $\times$ 4\carcsec{}7 large subfield containing two
small network patches with a dimension of $\approx$1\arcsec{} (see
\fig{fig1.eps}, position $x=$10\carcsec{}5, $y=$5\carcsec{}8 and
$x=$12\carcsec{}5, $y=$8\carcsec{}3). On average we find roughly 10 network
patches with similar Stokes signals per \imax{} snapshot.

The standard \imax{} data reduction routines were used to perform dark-current
subtraction, flat-field correction, and cross-talk removal. The noise level in
the Stokes $V$, $Q$ and $U$ signals was $\approx3\times10^{-3}$. A spatial
resolution of 0\carcsec{}15$-$0\carcsec{}18 in all Stokes parameters was
achieved by combining the on-board image stabilization system
\citep{gandorfer:10,berkefeld:10} with a phase-diversity-based post processing
technique \citep{martinezpillet:10a}.

\section{Setup of inversions}

The inversions of the Stokes vector observed with \imax{} are carried out
using the SPINOR code \citep{frutiger:00a,frutiger:thesis}. This code
numerically solves the radiative transfer equation (RTE) under the assumption
of local thermodynamic equilibrium and minimizes the difference between the
measured profile and the computed synthetic profile using a
response-function-based Levenberg--Marquardt algorithm. The limited number of
wavelength points of the \imax{} measurements requires the usage of a simple
model atmosphere with height-independent values for the magnetic field vector
and the line-of-sight velocity. This model atmosphere is in some ways
comparable to Milne-Eddington atmospheres that involve nine free parameters,
four of them being so-called ad hoc parameters without a one-to-one connection
with atmospheric parameters (i.e., the line center to continuum opacity ratio,
two parameters describing the linear source function, and damping). In
contrast, our numerical solution of the RTE requires only eight free
parameters: the magnetic field strength, $B$, inclination, $\gamma$, azimuth,
$\chi$, the line-of-sight velocity, $v_{\rm{LOS}}$, macro- and
micro-turbulence, $\xi_{\rm{mac}}$ and $\xi_{\rm{mic}}$, and two parameters
describing the temperature stratification. We used the temperature
stratification defined by the HSRASP model as the basis \citep{chapman:79} and
modified it by adding a linear function defined by an offset, $T_0$, and a
temperature gradient offset (changing the gradient in the HSRASP profile),
$T_{\rm{Grad}}$:
\begin{equation}
\label{temp}
T(\log \tau) ~=~ T_{\rm{HSRASP}}(\log \tau) + T_0 + T_{\rm{Grad}} \cdot \log\tau .
\end{equation}
Positive values of $T_{\rm{Grad}}$ lead to a steepening of the temperature
profile with respect to HSRASP while negative values lead to a flatter
profile.  To avoid a temperature increase for large negative values of
$T_{\rm{Grad}}$ in the upper atmospheric layers, we forced the gradient of
$T(\log\tau)$ to be steeper than 150\,K per unit of $\log\tau$, similar to the
gradient of the PLA atmosphere of \citet[see \sref{results}]{solanki:92c} at a
height of $\log\tau=-3$. As a consequence of the high spatial resolution of
the \imax{} observations, we set the macro-turbulence velocity to $0$.  For a
proper comparison with the measured \imax{} profiles, we convolve the
synthetic profiles with the filter curve, modeled with a Gaussian with a full
width at half-maximum of 85\,m\AA{} \citep{martinezpillet:10a}.

\begin{figure}
  \centering
  \includegraphics[width=\linewidth]{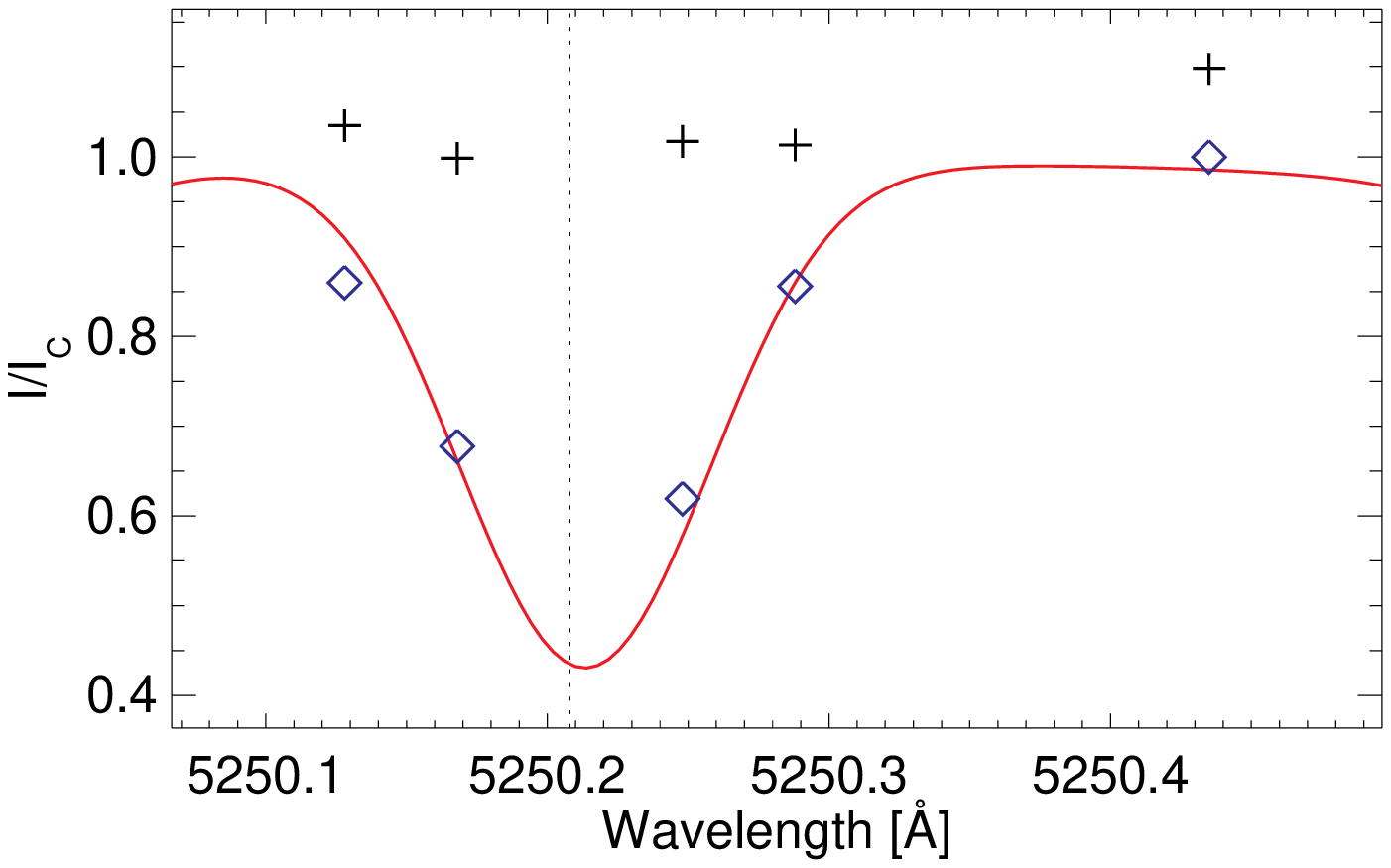}
  \caption{\label{fig2.eps}Stokes $I$ profiles. The diamonds represent the
    profile measured at the five \imax{} wavelength points averaged over the
    field of view observed with \imax{}. The solid line is a synthetic profile
    resulting from the HSRA model. The crosses correspond to the measured $I$
    profile in a network patch marked with (1) in \fig{fig1.eps}.}
\end{figure}

In \fig{fig2.eps}, the $I$ profile computed from the HSRA model is
overplotted on the profile averaged over the \imax{} field of view. This
average quiet-Sun profile (diamond symbols in \fig{fig2.eps}) is
consistent with a synthetic profile (solid red line) computed from the HSRA
model \citep{gingerich:71} assuming a micro- and macro-turbulence of
0.8\kms{} and 0\kms{}, respectively, and then convolved with a 85\,m\AA{}
Gaussian, corresponding to the bandpass of the \imax{} etalon. The good match
of the measured average Stokes $I$ profile with the synthetic profile computed
from the HSRA atmosphere justifies the usage of this atmospheric model as the
basis for the inversions. The temperature of the HSRA atmosphere at
$\log\tau=0$ is 6390\,K with a gradient at this height of 1900\,K per unit of
$\log\tau$. We normalize all Stokes profiles to the average quiet-Sun
continuum level. This normalization replaces the photometric calibration
necessary to use Planck's law for determining the temperatures in the
continuum forming layers.  The intensity in the line core constrains the
temperature in higher layers of the solar atmosphere. For later comparison
(see \sref{results}), we overplot with cross symbols the $I$ profile measured
in one of the network patches marked with (1) in \fig{fig1.eps}, normalized to
the average quiet-Sun continuum.

Neighboring spectral lines might be shifted to the \imax{} wavelength interval
when large line-of-sight velocities are present. In this analysis, we
therefore included the \coi{} line at 5250.0\,\AA{} and the \fei{} line at
5250.6\,\AA{} These lines are calculated assuming the same atmospheric model
as used for the synthesis of the \imax{} line (\fei{} 5250.2\,\AA{}).  The
decrease in intensity at both boundaries in the $I$ profiles shown in
\figs{fig2.eps} and \ref{fig4.eps} stems from the neighboring \coi{}
and \fei{} lines.

\section{Results}\label{results}

\begin{figure}
  \centering
  \includegraphics[width=\linewidth]{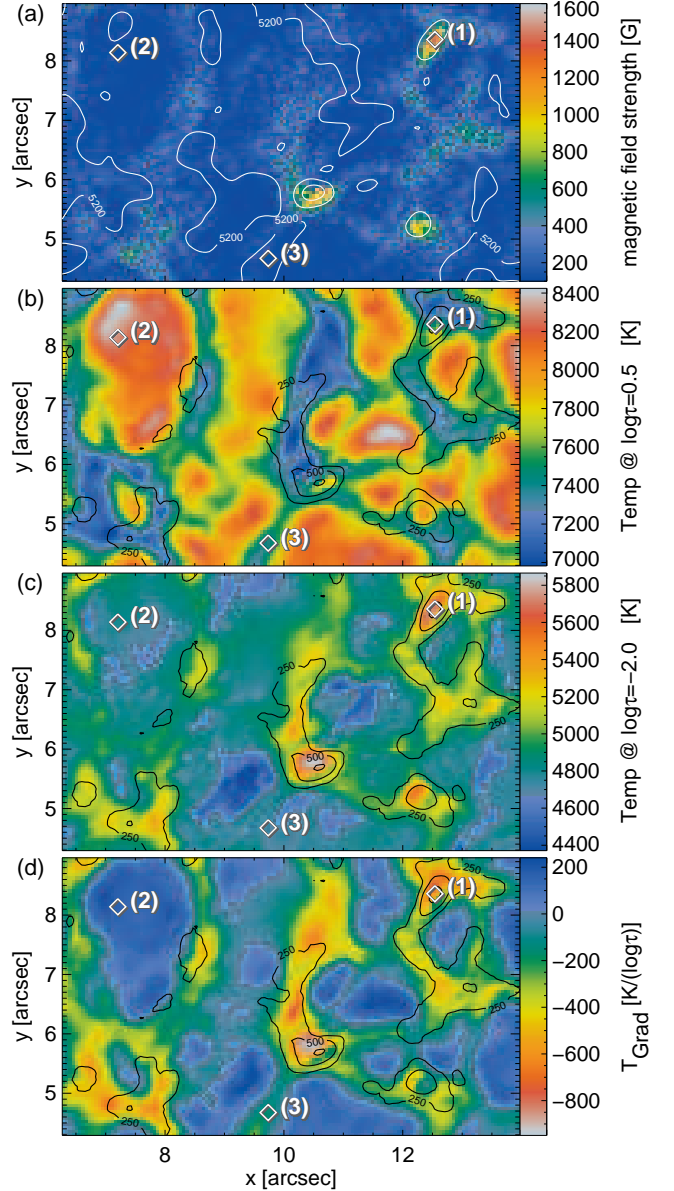}
  \caption{\label{fig3.eps}Maps of magnetic field strength (panel (a)),
    temperature at $\log\tau=0.5$ and $\log\tau=-2.0$ (panels (b) and (c)),
    and temperature gradient offset $T_{\rm{Grad}}$ as defined in \eqref{temp}
    (panel (d)). The contour lines of the magnetic field map (panel (a)) show
    the temperature at $\log\tau=-1.0$ (5200, 5500, and 5800\,K), the contour
    lines in the lower three panels show the height-independent magnetic field
    strength (250, 500, and 1000\,G).}
\end{figure}

The inversions were applied to an \imax{} set of Stokes parameter images. We
now discuss the results obtained in a box of 7\carcsec{}7 $\times$
4\carcsec{}7 (140 $\times$ 85 pixels), containing a region of enhanced Stokes
$V$ signal (see \fig{fig1.eps}). The resulting maps for the magnetic field
strength (constant with height), the temperature at $\log\tau=+0.5$ and
$-2.0$, and the temperature gradient offset $T_{\rm{Grad}}$ (see \eqref{temp})
are shown in \fig{fig3.eps}. Contour lines of the magnetic field strength
are overlaid on the temperature and temperature gradient maps, contour lines
of the temperature at $\log\tau=-1.0$ are plotted on top of the magnetic field
strength map.  The magnetic field outside the magnetic patches is weak and
lies close to or below the detection limit of \imax{}. In the center of the
magnetic patch (1) the field strength reaches values of up to 1.45\,kG. Note
that the inversions return these field strengths without introducing a
magnetic filling factor. This is one among many other examples already found
in the \imax{} data set.

The temperature map at $\log\tau=+0.5$ (panel (b) of \fig{fig3.eps})
clearly reflects the typical structure of the quiet-Sun in the deep layers of
the photosphere, with hot granules and cool intergranular lanes \citep[see,
e.g.,][]{borrero:02}. With increasing height, the temperature in the granules
falls off more rapidly than in the intergranular lanes, resulting in a
reversed granulation pattern \citep{cheung:07b}. Panel (c) of
\fig{fig3.eps} shows the temperature at $\log\tau=-2.0$, a height between
these extremes. At this height, hotter regions generally have stronger
magnetic fields, indicating a shallow temperature gradient in the magnetic
regions. This is reflected in the map of the temperature gradient offset,
$T_{\rm{Grad}}$, shown in panel (d) of \fig{fig3.eps}. In the granules the
temperature falls off more rapidly than in the HSRASP atmospheric model (blue
colors), whereas in the intergranular lanes, and especially in the magnetic
field patches, the gradient is significantly lower (negative values of
$T_{\rm{Grad}}$). As a result of this low gradient, the temperature in the
kilo-Gauss features at $\log\tau=-2.0$ is $\approx$1000\,K higher than in the
granules surrounding the magnetic patches.

\begin{figure}
  \centering
  \includegraphics[width=\linewidth]{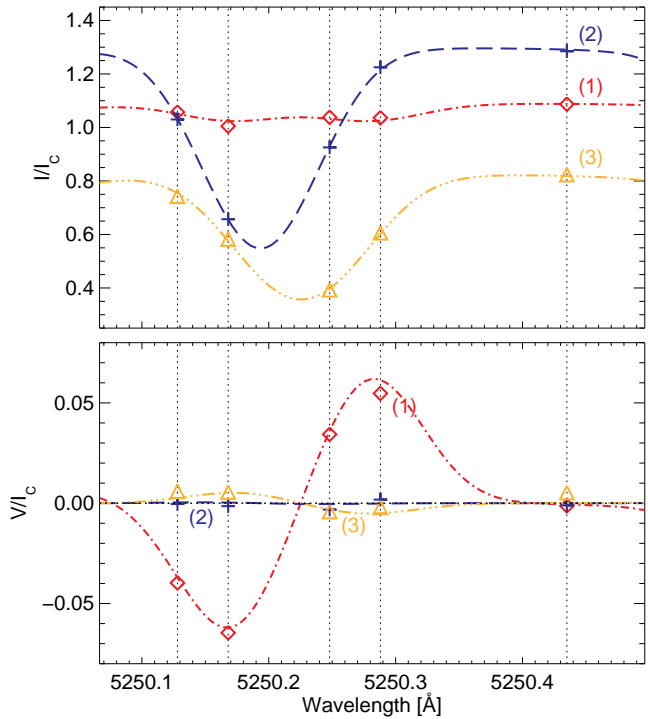}
  \caption{\label{fig4.eps}Measured (symbols) and fitted (lines) Stokes
    $I$ (top) and $V$ (bottom) profiles for the pixels (1), (2), and (3) in
    \fig{fig1.eps}. The vertical, dotted lines indicate the \imax{} filter
    positions. The decrease in continuum intensity at both boundaries and the
    increase in Stokes $V$ at the red boundary are caused by the neighboring
    \coi{} and \fei{} lines.}
\end{figure}

Measured (symbols) and fitted (lines) Stokes $I$ and $V$ profiles within a
network patch (pixel 1, red), a granule (pixel 2, blue), and an intergranular
lane (pixel 3, yellow) are shown in \fig{fig4.eps} (pixel numbers
$(764,809)$, $(667,805)$, and $(713,742)$ in the \imax{} data set
163--209). Owing to its low excitation potential of 0.12\,eV, the \fei{}
5250.2\,\AA{} line is highly sensitive to temperature. The high temperature at
the line formation height in the magnetic patch (1) therefore results in an
extremely weak Stokes $I$ profile, with a line depth of only 6\% of the local
continuum (see \fig{fig4.eps}).  At the same time, the continuum is
higher than the average quiet-Sun continuum.  The line depths of a typical
granular profile (2) and an intergranular profile (3) are significantly larger
($\approx$57\% and 56\%, respectively, of their continuum levels).

\begin{figure}
  \centering
  \includegraphics[width=\linewidth]{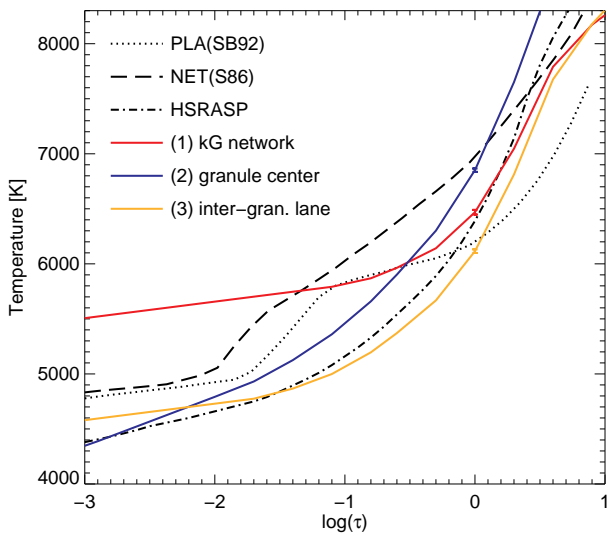}
  \caption{\label{fig5.eps}Temperature profile for model atmospheres and as
  retrieved from the inversions for the pixels (1), (2), and (3) in
  \fig{fig1.eps}.}
\end{figure}


In \fig{fig5.eps}, we compare the temperature stratifications returned
for the three selected pixels featured in \fig{fig4.eps}. The
temperature gradient in the intergranular lane (3) is somewhat flatter than of
the HSRASP atmosphere. The temperature in the center of the granule (2) is
higher in deep layers and falls off rapidly with height. The kilo-Gauss
network patch (1) shows high temperatures that, for the most part, lie between
the empirical network \citep[labeled NET(S86)]{solanki:86} and the plage
models \citep[PLA(SB92)]{solanki:92c}. In the upper layers it is even hotter
than both of these atmospheres, but not well constrained. Similar temperature
stratifications (and also field strengths) were obtained from the analysis of
spectropolarimetric observations in the \fei{} 630\,nm line pair \citep[see,
e.g.,][]{sanchezalmeida:00}, but now, for the first time, we obtain this
temperature stratification without introducing a magnetic filling factor.  The
relevant atmospheric parameters resulting from the inversion for pixel (1) are
$T_0=6467$\,K, $T_{\rm{Grad}}=-652$\,K/$\log\tau$, $B=1447$\,G,
$\gamma=169^\circ$, and $\xi_{\rm{mic}}=0.1$\kms{}, for pixel (2)
$T_0=6865$\,K, $T_{\rm{Grad}}=167$\,K/$\log\tau$, and
$\xi_{\rm{mic}}=1.4$\kms{}, and for pixel (3) $T_0=6139$\,K,
$T_{\rm{Grad}}$=$-168$\,K/$\log\tau$, and $\xi_{\rm{mic}}=1.7$\kms{}.  The
inclination of the magnetic field for pixel (1) is likely to be more vertical
to the solar surface since a small linear polarization signal could not be
well constrained by the inversion. The kilo-Gauss nature of the magnetic patch
persists even under the assumption of a completely vertical ($180^\circ$)
field. The magnetic field strength and inclination for pixels (2) and (3)
cannot be determined reliably due to the low polarization signal.

\section{Summary and Conclusions}

The \imax{} instrument on board the \sunrise{} mission allowed photospheric
magnetic fields to be analyzed with unprecedented resolution.  The quiet-Sun
region studied in this Letter contains two network patches with the
characteristic observational signature of an extremely weak Stokes $I$
profile, associated with a relatively strong signal in Stokes $V$.

Without requiring a magnetic filling factor, the inversion of the Stokes
profiles of a single pixel within such a network patch returns a magnetic
field strength of 1.45\,kG. This value is consistent with field strengths
determined from the \fei{} 5250.2\,\AA{} line in the quiet-Sun network using
indirect techniques. Thus, \citet{stenflo:85} find values of
$800$--$1100$\,G. Similarly, \citet{grossmanndoerth:96} obtained a
distribution of intrinsic field strengths in the quiet-Sun from this line,
which peak around $800$--$1000$\,G.



The extremely weak Stokes $I$ profiles in the network patches point to their
high temperature in the higher atmospheric layers. We find the temperature at
heights of $\log\tau=-1$, corresponding to the formation height of the line
core in these areas, to lie significantly above the temperature in the
surrounding area. The inversions directly give rise to an atmospheric
stratification of physical quantities consistent with existing flux tube
models in network plage regions \citep{solanki:86,solanki:92c}, and with the
properties of magnetic flux concentrations in magnetohydrodynamic simulations
\citep{voegler:05a}.

We can identify 5--10 similar network patches within every \imax{}
snapshot. Inversions of these network patches yield field strengths of up to
1.8\,kG. The temperature stratification and the other atmospheric parameters
are similar to the values presented in \fig{fig5.eps} and in
\sref{results} for pixel (1). Different from the analysis of
\citet{berger:04}, these kilo-Gauss structures are no obvious flux
concentrations seen as micropores in the intensity images.

Both results, the kilo-Gauss field strength and the high temperatures in the
mid- to upper photosphere, lead us to the conclusion that the observed network
patch is a magnetic flux tube\footnote{The rapid expansion of a vertical flux
  tube with height results in a strong magnetic field gradient, which was not
  included in this analysis. Test inversions including a magnetic field
  gradient of -3\,G\,km$^{-1}$ \citep{martinezpillet:97} reproduce the
  observed Stokes profiles equally well as inversions with a
  height-independent magnetic field strength. This gradient model yields field
  strengths at the formation height of the line core ($\log\tau\approx\-1$)
  very similar to the field strengths presented in this Letter.}. The
unprecedented spatial resolution of 0\carcsec{}15$-$0\carcsec{}18, achieved
with \imax{} on \sunrise{}, was sufficient to resolve this flux tube. This has
been a long cherished aim in solar physics. We have demonstrated that with a
1\,m class telescope under ideal conditions it is achievable, even in the
quiet-Sun.  The data set obtained during the \sunrise{} flight will allow us
to analyze the temporal evolution of these flux tubes in the quiet-Sun as well
as to study their properties in a statistical sense without dilution due to
serious intermingling with light from field-free gas. Additionally, the
connection between the shallow temperature gradient in these flux tubes and
the bright points at chromospheric heights observed with the \sunrise{} Filter
Imager \citep[\sufi{};][]{gandorfer:10,hirzberger:10a} poses an interesting
topic for a future analysis.

\begin{acknowledgements}
  The German contribution to \sunrise{} is funded by the Bundesministerium
  f\"{u}r Wirtschaft und Technologie through Deutsches Zentrum f\"{u}r Luft-
  und Raumfahrt e.V. (DLR), Grant No. 50~OU~0401, and by the Innovationsfond
  of the President of the Max Planck Society (MPG). The Spanish contribution
  has been funded by the Spanish MICINN under projects ESP2006-13030-C06 and
  AYA2009-14105-C06 (including European FEDER funds). The HAO contribution was
  partly funded through NASA grant NNX08AH38G. This work has been partly
  supported by the WCU grant (No R31-10016) funded by the Korean Ministry of
  Education, Science and Technology.
\end{acknowledgements}

\end{document}